\documentclass{egpubl}
\usepackage{pg2024}

\SpecialIssueSubmission    

\usepackage[utf8]{inputenc}									
\usepackage[T1]{fontenc}									
\usepackage{dfadobe}

\usepackage{cite}  
\BibtexOrBiblatex
\electronicVersion
\PrintedOrElectronic
\ifpdf \usepackage[pdftex]{graphicx} \pdfcompresslevel=9
\else \usepackage[dvips]{graphicx} \fi

\usepackage{egweblnk} 

\title[Real-Time Glints and Area Lights]%
      {Real-Time Rendering of Glints in the Presence of Area Lights}

\newif\ifanonymous

\ifanonymous
\author[paper1295]
{\parbox{\textwidth}{\centering
		paper1295
	} \\
	{\parbox{\textwidth}{\centering }}
}
\else
\author[Tom Kneiphof \& Reinhard Klein]
{\parbox{\textwidth}{\centering
		Tom Kneiphof$^{1}$\orcid{0000-0002-7237-2431}
		and Reinhard Klein$^{1}$\orcid{0000-0002-5505-9347}
	} \\
	{\parbox{\textwidth}{\centering $^1$University of Bonn, Germany}}
}
\fi

\usepackage{amsmath}    
\usepackage{amssymb}    
\usepackage{mathtools}
\usepackage{bm}         
\usepackage{cancel}     
\usepackage{nicefrac}   

\usepackage{siunitx}    
\usepackage{csquotes}

\usepackage{subcaption}
\usepackage{cleveref}
\crefname{equation}{Eq.}{Eqs.}
\crefname{figure}{Fig.}{Figs.}
\crefname{table}{Tab.}{Tabs.}
\crefname{section}{Sec.}{Secs.}
\crefname{appendix}{Appendix}{Appendices}

\usepackage{tikz}
\usetikzlibrary{positioning,calc}
\usetikzlibrary{shapes.misc}
\usetikzlibrary{arrows.meta}
\usetikzlibrary{patterns}
\usetikzlibrary{decorations.pathreplacing,calligraphy}
\tikzset{cross/.style={cross out, draw=black, minimum size=2*(#1-\pgflinewidth), inner sep=0pt, outer sep=0pt},cross/.default={1pt}}
\pgfdeclarelayer{bg}    
\pgfsetlayers{bg,main}  

\usepackage{pgfplots}
\usepackage{pgfplotstable}
\usepgfplotslibrary{polar}

\usepackage[l3]{csvsimple}

\usepackage[toc]{appendix}

\usepackage{booktabs}

\usepackage{url}
\newcommand{\hide}[1]{}

\definecolor{revisioncolor}{RGB}{34,139,34}
\newcommand{\revised}[1]{\textcolor{revisioncolor}{{}\ifx&#1&[\ldots]\else#1\fi{}}}

	{} 

\newcommand{\citeq}[1]{\citeleft\citen{#1}\citeright}

\let\originalleft\left
\let\originalright\right
\renewcommand{\left}{\mathopen{}\mathclose\bgroup\originalleft}
\renewcommand{\right}{\aftergroup\egroup\originalright}

\newcommand{\tikzboxed}[2][]{\begin{tikzpicture}[baseline=(current bounding box.base)]\node[anchor=base,draw,#1]{$\displaystyle #2$};\end{tikzpicture}}

\newcommand{\dx}[1]{\;\mathrm{d}#1}
\newcommand{\sprod}[2]{\langle #1 \,, #2 \rangle}
\newcommand{\derivative}[2]{{ \frac{\partial#1}{\partial#2} }}
\newcommand{\nicederivative}[2]{{ \nicefrac{\partial#1}{\partial#2} }}

\newcommand{\Ex}[2][]{\mathbb{E}_{#1}\left[#2\right]}

\newcommand{\srof}[1]{{|#1|}}

\newcommand{\dirH}{{\mathbf{h}}}
\newcommand{\dirM}{{\mathbf{m}}}
\newcommand{\dirN}{{\mathbf{n}}}

\newcommand{\dirI}{{\omega_i}}
\newcommand{\dirO}{{\omega_o}}
\newcommand{\dirL}{{\omega_i}}

\newcommand{\dirV}{{\omega_o}}

\newcommand{\dirHk}[1][k]{{\dirH^{(#1)}}}

\newcommand{\domainIsup}[1][]{{\ifx&#1&\Omega_i\else\Omega_i^{#1}\fi}}
\newcommand{\domainI}{{\domainIsup}}
\newcommand{\domainHsup}[1][]{{\ifx&#1&\Omega_h\else\Omega_h^{#1}\fi}}
\newcommand{\domainH}{{\domainHsup}}
\newcommand{\domainL}{{\Omega_i}}
\newcommand{\domainHemi}{{\mathcal{H}}}

\newcommand{\hdotl}{{ \sprod{\dirH}{\dirL} }}

\newcommand{\ndotl}{{ \sprod{\dirN}{\dirL} }}
\newcommand{\ndotv}{{ \sprod{\dirN}{\dirV} }}

\newcommand{\pxFoot}{{\mathcal{P}}}

\newcommand{\brdfSymb}[1][]{{#1\rho}}
\newcommand{\brdfArgs}[2][]{{\brdfSymb[#1]\left(#2\right)}}
\newcommand{\brdf}[1][]{{\brdfArgs[#1]{\dirI,\dirO}}}

\newcommand{\brdfnolSymb}[1][]{{\brdfSymb[\tilde]}}
\newcommand{\brdfnolArgs}[2][]{{\brdfnolSymb[#1]\left(#2\right)}}
\newcommand{\brdfnol}[1][]{{\brdfnolArgs[#1]{\dirI,\dirO}}}

\newcommand{\fresnelSymb}[1][]{{#1F}}
\newcommand{\fresnelArgs}[2][]{{\fresnelSymb[#1]\left(#2\right)}}
\newcommand{\fresnel}[1][]{{\fresnelArgs[#1]{\dirI, \dirH}}}

\newcommand{\geomSymb}[1][]{{#1G}}
\newcommand{\geomArgs}[2][]{{\geomSymb[#1]\left(#2\right)}}
\newcommand{\geom}[1][]{{\geomArgs[#1]{\dirI, \dirO}}}

\newcommand{\ndfSymb}[1][]{{#1D}}
\newcommand{\ndfArgs}[2][]{{\ndfSymb[#1]\left(#2\right)}}
\newcommand{\ndf}[1][]{{\mathop{\ndfArgs[#1]{\dirH}}}}

\newcommand{\ndfFootSymb}{{\ndfSymb[\hat]_\pxFoot}}
\newcommand{\ndfFootArgs}[1]{{\ndfFootSymb\left(#1\right)}}
\newcommand{\ndfFoot}{{\ndfFootArgs{\dirH}}}

\newcommand{\LiSymb}[1][]{{{#1L}_i}}
\newcommand{\LiArgs}[2][]{{\LiSymb[#1]\left(#2\right)}}
\newcommand{\Li}[1][]{{\LiArgs[#1]{\dirI}}}

\newcommand{\totalLi}{{\LiSymb[]}}

\newcommand{\LoSymb}[1][]{{{#1L}_o}}
\newcommand{\LoArgs}[2][]{{\LoSymb[#1]\left(#2\right)}}
\newcommand{\Lo}[1][]{{\LoArgs[#1]{\dirO}}}

\newcommand{\LoFootSymb}{{\hat{L}_{o,\pxFoot}}}
\newcommand{\LoFootArgs}[1]{{\LoFootSymb\left(#1\right)}}
\newcommand{\LoFoot}{{\LoFootArgs{\dirO}}}

\newcommand{\Nfoot}{{N_\pxFoot}}
\newcommand{\probHsup}[1][]{{\ifx&#1&p_{\domainH}\else p_{\domainH}^{#1}\fi}}
\newcommand{\probH}{{\probHsup}}
\newcommand{\intDLong}[1][]{{ \int_{\domainHsup[#1]} \ndf \dx{\dirH} }}
\newcommand{\intDHemiLong}{{ \int_\domainHemi \ndf \dx{\dirH} }}
\newcommand{\intD}[1][]{{ \mathcal{D}_{\domainHsup[#1]} }}
\newcommand{\intDHemi}{{ \mathcal{D}_\domainHemi }}

\newcommand{\FGD}{{\mathrm{FGD}}}
\newcommand{\ndfFGD}{{\mathcal{D}_{\mathrm{PR}}}}

\newcommand{\meanRefl}{{\mathcal{FG}}}

\newcommand{\approxbutequal}{\approx} 

\definecolor{myred}{RGB}{228,26,28}

\definecolor{set1red}{RGB}{238,157,3}
\definecolor{set1green}{RGB}{154,209,1}
\definecolor{set1blue}{RGB}{6,140,209}

\begin{document}

\newcommand{\imgpathteaser}[1]{images/beethoven/scale4_hr/Beethoven\the\numexpr#1-2\relax.png}
\newcommand{\imgtrimteaser}{16}
\def\scaleteaser{2}
\newcommand{\imgtriminsetteaser}[1]{\numexpr#1*\scaleteaser\relax}
\teaser{
	\centering
	%
	%
	\tikzset{
		insetAZoom/.style={
			draw, set1green, line width=3.5pt, line join=round
		},
		insetBZoom/.style={
			draw, set1blue, line width=3.5pt, line join=round
		},
		insetAOverlay/.style={
			draw, set1green, line width=1pt, line join=round
		},
		insetBOverlay/.style={
			draw, set1blue, line width=1pt, line join=round
		}
	}
	\subfloat[Low Density]{%
		\begin{tikzpicture}[node distance=.5em,every node/.style={inner sep=0,outer sep=0}]
			\node[anchor=north west] (main) at (0,0) {\includegraphics[trim={\imgtriminsetteaser{16} 0 \imgtriminsetteaser{16} 0},clip,width=0.275\linewidth]{\imgpathteaser{10}}};
			\begin{scope}[x={(main.north east)},y={(main.south west)}, shift={(main.north west)}, xscale=0.016447368,yscale=0.01953125]
				\draw[insetAOverlay] (22.5,28.8) rectangle +(4.8,+4.8);
				\draw[insetBOverlay] (36.9,28.9) rectangle +(4.8,+4.8);
			\end{scope}
		\end{tikzpicture}%
		\label{fig:teaser:low}%
		\hspace{1pt}%
	}
    \subfloat[Medium Density]{%
    	\hspace{1pt}%
    	\begin{tikzpicture}[node distance=.5em,every node/.style={inner sep=0,outer sep=0}]
			\node[anchor=north west] (main) at (0,0) {\includegraphics[trim={\imgtriminsetteaser{16} 0 \imgtriminsetteaser{16} 0},clip,width=0.275\linewidth]{\imgpathteaser{13}}};
			\begin{scope}[x={(main.north east)},y={(main.south west)}, shift={(main.north west)}, xscale=0.016447368,yscale=0.01953125]
				\draw[insetAOverlay] (22.5,28.8) rectangle +(4.8,+4.8);
				\draw[insetBOverlay] (36.9,28.9) rectangle +(4.8,+4.8);
			\end{scope}
		\end{tikzpicture}%
		\label{fig:teaser:medium}%
		\hspace{1pt}%
	}
	\subfloat[High Density]{%
		\hspace{1pt}%
		\begin{tikzpicture}[node distance=.5em,every node/.style={inner sep=0,outer sep=0},remember picture]
			\node[anchor=north west] (main) at (0,0) {\includegraphics[trim={\imgtriminsetteaser{16} 0 \imgtriminsetteaser{16} 0},clip,width=0.275\linewidth]{\imgpathteaser{17}}};
			\begin{scope}[x={(main.north east)},y={(main.south west)}, shift={(main.north west)}, xscale=0.016447368,yscale=0.01953125]
				\draw[insetAOverlay] (22.5,28.8) rectangle +(4.8,+4.8);
				\draw[insetBOverlay] (36.9,28.9) rectangle +(4.8,+4.8);
			\end{scope}
			
		\end{tikzpicture}%
		\label{fig:teaser:high}%
		\hspace{2pt}%
	}
	\subfloat[Insets]{%
		\hspace{3pt}%
		\hspace{0.068\linewidth}%
		\hspace{5pt}%
		\hspace{0.068\linewidth}%
		\hspace{3pt}%
		\begin{tikzpicture}[node distance=5pt,every node/.style={inner sep=0,outer sep=0},remember picture,overlay]
			\node [insetAZoom, right=5pt of main] (mid_a) {\includegraphics[trim={\imgtriminsetteaser{241} \imgtriminsetteaser{176} \imgtriminsetteaser{351} \imgtriminsetteaser{288}},clip,width=0.068\linewidth]{\imgpathteaser{13}}};
			\node [insetBZoom, right=of mid_a] (mid_b) {\includegraphics[trim={\imgtriminsetteaser{385} \imgtriminsetteaser{175} \imgtriminsetteaser{207} \imgtriminsetteaser{289}},clip,width=0.068\linewidth]{\imgpathteaser{13}}};
			
			\node [insetAZoom, anchor=north west,above=of mid_a] (low_a) {\includegraphics[trim={\imgtriminsetteaser{241} \imgtriminsetteaser{176} \imgtriminsetteaser{351} \imgtriminsetteaser{288}},clip,width=0.068\linewidth]{\imgpathteaser{10}}};
			\node [insetBZoom, right=of low_a] (low_b) {\includegraphics[trim={\imgtriminsetteaser{385} \imgtriminsetteaser{175} \imgtriminsetteaser{207} \imgtriminsetteaser{289}},clip,width=0.068\linewidth]{\imgpathteaser{10}}};
			
			\node [insetAZoom, below=of mid_a] (high_a) {\includegraphics[trim={\imgtriminsetteaser{241} \imgtriminsetteaser{176} \imgtriminsetteaser{351} \imgtriminsetteaser{288}},clip,width=0.068\linewidth]{\imgpathteaser{17}}};
			\node [insetBZoom, right=of high_a] (high_b) {\includegraphics[trim={\imgtriminsetteaser{385} \imgtriminsetteaser{175} \imgtriminsetteaser{207} \imgtriminsetteaser{289}},clip,width=0.068\linewidth]{\imgpathteaser{17}}};
			
		\end{tikzpicture}%
		\label{fig:teaser:insets}%
	}
	\caption{
		We compute the appearance of a glittery surface illuminated by two area light sources with equal overall intensity but different size in real time.
		(\subref{fig:teaser:low}) When the glints are sparsely distributed across the surface, fewer microfacets are oriented to reflect light from the small light than from the large light.
		%
		(\subref{fig:teaser:medium}) The small light induces more pronounced glints than the large light for medium glint densities.
		(\subref{fig:teaser:high}) Our approach converges to the continuous microfacet BRDF for large glint densities.
		The insets in (\subref{fig:teaser:low}-\subref{fig:teaser:high}) are shown from top to bottom in (\subref{fig:teaser:insets}).
	}
	\label{fig:teaser}
}

\maketitle

\begin{abstract}
	Many real-world materials are characterized by a glittery appearance.
	%
	Reproducing this effect in physically based renderings is a challenging problem due to its discrete nature, especially in real-time applications which require a consistently low runtime.
	Recent work focuses on glittery appearance illuminated by infinitesimally small light sources only. 
	For light sources like the sun this approximation is a reasonable choice.
	%
	In the real world however, all light sources are fundamentally area light sources.
	%
	%
	In this paper, we derive an efficient method for rendering glints illuminated by spatially constant diffuse area lights in real time.
	To this end, we require an adequate estimate for the probability of a single microfacet to be correctly oriented for reflection from the source to the observer.
	A good estimate is achieved either using linearly transformed cosines (LTC) for large light sources, or a locally constant approximation of the normal distribution for small spherical caps of light directions.
	To compute the resulting number of reflecting microfacets, we employ a counting model based on the binomial distribution.
	In the evaluation, we demonstrate the visual accuracy of our approach, which is easily integrated into existing real-time rendering frameworks, especially if they already implement shading for area lights using LTCs and a counting model for glint shading under point and directional illumination.
	Besides the overhead of the preexisting constituents, our method adds little to no additional overhead.
	%


\end{abstract}

%
\section{Introduction}
\label{sec:introduction}
\begin{figure}
	\centering
	\subfloat[Setup]{%
		\includegraphics[height=2.9cm,trim={360 217 344 375},clip]{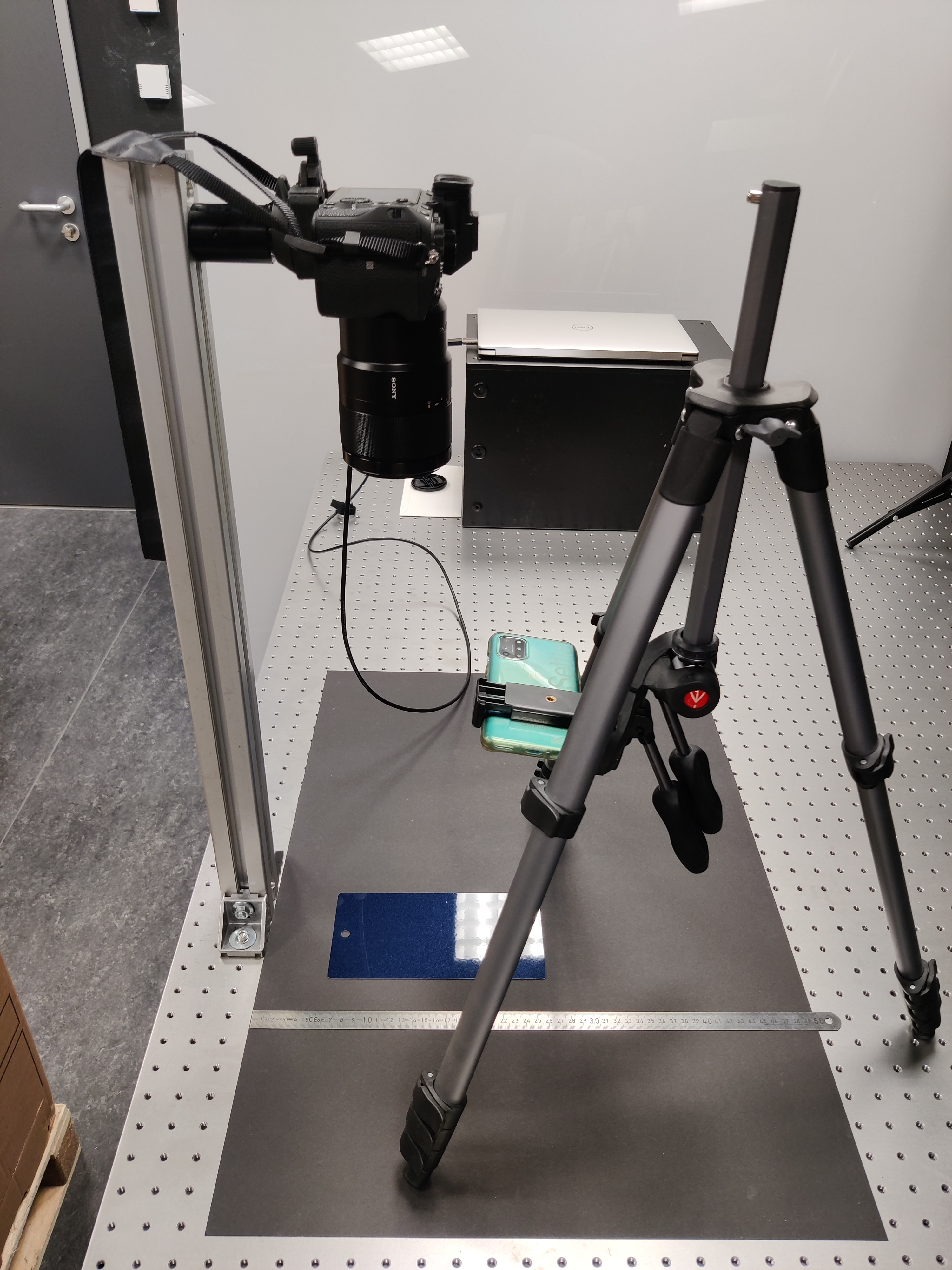}%
		\hspace{2pt}%
		\label{fig:photo_flakes:setup}%
	}
	\tikzset{
		inset/.style={
			draw,set1green,line width=2pt,line join=round
		},
		highlight/.style={
			set1green,line width=0.666pt,line join=round
		}
	}
	\subfloat[$r=\SI{10}{\milli\meter}$]{%
		\hspace{1pt}%
		\begin{tikzpicture}[node distance=.2em,every node/.style={inner sep=0,outer sep=0}]
			\node[anchor=north west] (main) at (0,0) {\includegraphics[height=2.9cm]{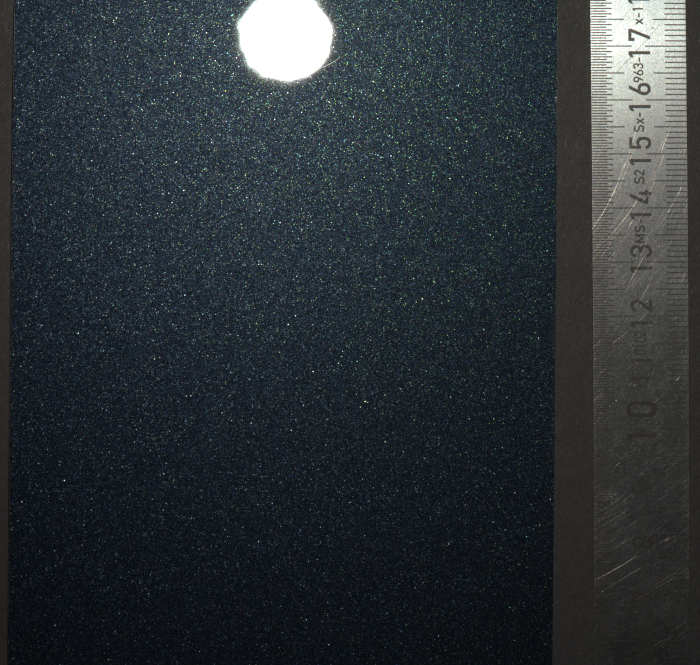}};		\begin{scope}[x={(main.north east)},y={(main.south west)}, shift={(main.north west)}, xscale=0.014285714,yscale=0.015037594]
				\draw[highlight] (24.1,18.5) rectangle +(6.4,+6.4);
			\end{scope}
			\node[draw, above left=of main.south east,inset] {\includegraphics[height=1cm, trim=241 416 395 185,clip]{images/photo_flakes/out_r1.0_f8_i1600_full_4x.png}};
		\end{tikzpicture}%
		\hspace{2pt}%
		\label{fig:photo_flakes:10}%
	}
	\subfloat[$r=\SI{35}{\milli\meter}$]{%
		\hspace{2pt}%
		\begin{tikzpicture}[node distance=.2em,every node/.style={inner sep=0,outer sep=0}]
			\node[anchor=north west] (img) at (0, 0) {\includegraphics[height=2.9cm]{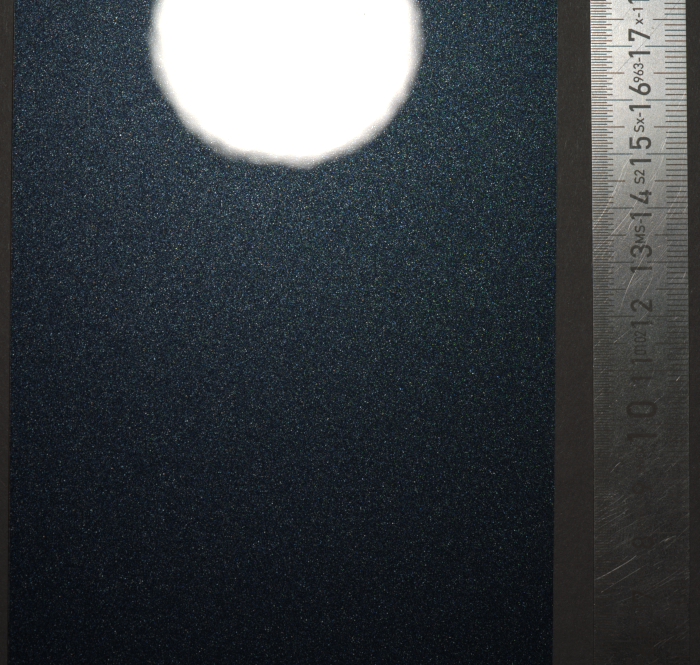}};		\begin{scope}[x={(main.north east)},y={(main.south west)}, shift={(main.north west)}, xscale=0.014285714,yscale=0.015037594]
				\draw[highlight] (24.1,18.5) rectangle +(6.4,+6.4);
			\end{scope}
			\node[above right=of main.south west,inset] {\includegraphics[height=1cm, trim=241 416 395 185,clip]{images/photo_flakes/out_r3.5_f8_i1600_full_4x.png}};
		\end{tikzpicture}%
		\label{fig:photo_flakes:35}%
	}
	\caption{
		The effect of real-world area lights is shown on a sample of car paint with metallic flakes.
		(\subref{fig:photo_flakes:setup}) The camera and a smartphone display are placed $\SI{50}{\centi\meter}$ and $\SI{30}{\centi\meter}$ above the sample.
		We took two exposure series illuminated by a small (\subref{fig:photo_flakes:10}) and a large (\subref{fig:photo_flakes:35}) light source, respectively.
		The shapes of the light sources have been realized using the display of a smartphone, and the reconstructed HDR images have been adjusted in post-processing such that the light sources emit an equal amount of radiant power.
		The experiment shows that the smaller light source results in sparser but brighter glints, and the larger light source produces a larger number of weaker glints.
	}
	\label{fig:photo_flakes}
\end{figure}

The real world contains a vast number of materials with intricate light reflection properties.
Computer graphics researchers have been on their heels to find suitable digital representations that are cost efficient and simultaneously produce accurate physically based renderings for decades \cite{smits1992newton}.
In particular real-time applications have the additional constraint that a full image has to be created consistently every few milliseconds on cheap consumer hardware.
Therefore, a full simulation of the light transport is often not feasible, requiring low-cost approximations that still produce results close to the ground truth.
A common category of such complex materials are formed by those exhibiting glittery appearance, ranging from metallic car paints over rough plastic/metal to literal glitter, all kinds of decorative ornamentation and even granular substances like sand.
Such a glittery appearance is characterized by individual highlights, that appear to randomly pop in and out of existence across the surface when the lighting or viewing configuration changes.
The glint effect produced by these kinds of materials is strongest under sharp uni-directional illumination, e.g. from the sun or a flashlight.
For larger area light sources like studio lights, the glints are much less pronounced but are usually still observable and produce some kind of grainy appearance.
We demonstrate this dependence of real-world glint intensity on the light source size in \cref{fig:photo_flakes}.
Most recent work in real-time glint rendering \cite{chermain2020procedural,deliot2023real} focuses on light sources that are infinitesimally small, i.e. directional or point lights.
This is a limitation on the achievable realism.
Even for very small light sources, their actual size can still influence the resulting appearance of a glittery surface.
The state of the art method for handling the interaction between microfacet surfaces and area light sources in real-time applications are linearly transformed cosines (LTC) \cite{heitz2016real}.
The goal of this work is to bridge the gap between glint rendering and area light sources in real-time applications.
%

%
Our main contribution lies in the physically motivated computation of the discrete probability that a microfacet is correctly oriented for a reflection from an area light source towards the observer, which has only been achieved for infinitesimally small light sources in the previous real-time glint rendering method \cite{deliot2023real} that we build upon.
For area light sources this probability is the result of an integration of the continuous normal distribution function (NDF), which we approximate using LTCs \cite{heitz2016real}.
This probability is then fed into the previous real-time glint rendering method \cite{deliot2023real}, which we use for evaluating the binomial distribution for a given number of microfacets inside of a pixel footprint and our discrete probability of correct orientation.
Given existing implementations for LTC integration for area light illumination \cite{heitz2016real} and real-time glint rendering \cite{deliot2023real}, our method treats these two approaches as black boxes and combines them together in a straightforward way.
This leads to a real-time method that supports realistic rendering of glittery appearance under area light illumination.
%
Moreover, our formulation allows for controlling the glittery appearance under standard infinitesimally small directional or point lights using a parameter that specifies the virtual size of the light source.
%


%
\section{Related Work}
\label{sec:related_work}
In the following, we review previous work on glint rendering, as well as real-time area lights.
Zhu et al. \cite{zhu2022recent} provide an overview over recent offline methods for glint rendering.
Most approaches can be classified as being stochastic \cite{jakob2014discrete}, or using high-resolution normal maps \cite{yan2014rendering}.
The body of work around real-time area lights largely builds around the introduction of LTCs into the graphics community \cite{heitz2016real}.
\paragraph{Stochastic Glints}
Our real-time approach relates to the stochastic glint category, which was pioneered by \cite{jakob2014discrete} for offline rendering, where discrete microfacets are randomly distributed in the spatio-angular domain.
Since then, several improvements have been made to that approach \cite{atanasov2016practical, wang2018fast}.
This line of work has been pushed over the real-time boundary using a prefiltering approach, that handles point and environment illumination \cite{wang2020real}.
%
%
In contrast, we replace the spatio-angular search for discrete microfacets with a counting model based on an approximation of binomial distributions which is naturally fast enough for real-time applications \cite{zirr2016real,deliot2023real}.
While it is desirable to importance-sample the glints\cite{chermain2021importance}, no importance sampling scheme currently exists for methods based on binomial distributions.
%

%

%
\paragraph{High Resolution Normal Maps}
Alternatively, a high resolution normal map can be employed to yield a glittery appearance \cite{yan2014rendering, yan2016position}.
This approach is sped up by a more efficient integration of the NDF over pixel footprint and light source \cite{gamboa2018scalable, atanasov2021multiscale, deng2022constant}, and the storage cost is reduced using procedural normal maps \cite{wang2020example}.
Neural and differentiable rendering techniques have also been employed to synthesize glittery appearance in this context\cite{kuznetsov2019learning,fan2022efficient}.
If the geometric features in the normal map become small enough, chromatic variations manifest due to diffraction\cite{yan2018rendering}.
The subjective optical speckle due to diffraction on random rough surfaces was investigated in more detail by Steinberg and Yan\cite{steinberg2022rendering}.
Using strong normal maps might cause energy loss due to back-facing normals \cite{schussler2017microfacet}, which was addressed by taking multiple-scattering into account\cite{chermain2019glint}.
In our case the multiple-scattering compensation for microfacet models commonly used in real-time applications is sufficient \cite{turquin2019practical}.
Scratches can be interpreted as a special kind of glints, which stretch across multiple pixels in screen space.
They are best handled using specialized methods\cite{raymond2016multi,werner2017scratch,velinov2018real}, but are compatible with high resolution normal maps.
Granular media like sand and snow also produce a glint-like appearance.
With granular media, a different problem of long scattering paths is relevant, which complements the glints produced by single-scattering reflections \cite{moon2007rendering, meng2015multi, muller2016efficient, zingsheim2024learning}.
\paragraph{Real-Time Glints}
Early procedural methods for real-time rendering produce highlights by intersecting the surface with a world-space grid\cite{bowles2015sparkly, wang2016robust}.
Multiple real-time approaches fall into the discrete stochastic line of work \cite{wang2020real,chermain2020procedural,zirr2016real,deliot2023real}, while other fast approaches employ normal map pre-filtering \cite{chermain2021real,tan2022real}.
\paragraph{Car Paints}
Metallic car paints are well known for their sparkling appearance, caused by metallic flakes embedded in the substrate\cite{ershov1999simulation,ershov2001rendering}.
Domain-specific data-driven representations have been developed for flakes\cite{rump2008photo, rump2009efficient,golla2017efficient}.
Another effect often observed in metallic paints are iridescent color shifts\cite{belcour2017practical,kneiphof2019real,kneiphof2022real}.
Guo et al. \cite{guo2018physically} introduced an offline method that merges both effects.
\paragraph{Real-Time Area Lights}
Linearly transformed cosines (LTC) have first been introduced for isotropic microfacet distributions under polygonal and line lights\cite{heitz2016real, heitz2017linear}.
A more elaborate fitting process is required to reduce artifacts for anisotropic microfacet distributions \cite{kt2022bringing}.
More complex linearly transformed distributions like spherical harmonics have also been tried\cite{allmenroder2021linearly}.
The emergence of real-time ray tracing prompted the necessity of improved importance sampling techniques to cast shadow rays \cite{dupuy2017spherical, peters2021poly, peters2021linear}.
Analytic area light shading can be efficiently combined with a stochastic estimate of the light source visibility \cite{heitz2018combining}.
However, soft shadows from area lights can also be estimated analytically by clipping the spherical polygon of the light source with that of an occluder \cite{kt2021fast}.
%
%
In the context of glint rendering it is crucial to account for the change in size of area lights within the penumbra, since the size of the light source affects the apparent glint pattern.
Liu et al.\cite{liu2023real} render iridescent color shifts from real-time area lights, while we add support for glints.
%

%
\section{Discrete Stochastic Microfacet Theory}
\label{sec:discrete_microfacet_theory}
This section introduces the discrete microfacet model that constitutes the theoretic foundation of our approach.
The core insight underlying our method was already observed by Jakob et al. \cite{jakob2014discrete} a decade ago:
We have to count the number of microfacets that are oriented in such a way, that they reflect light from the source to the observer.
The implications for real-time area light sources and infinitesimal light sources are subsequently presented in \cref{sec:area_lights} and \cref{sec:infinitesimal_lights}, respectively.
\paragraph{BRDF}
We start by defining the (continuous) microfacet bidirectional reflectance distribution function (BRDF)\cite{cook1982reflectance,walter2007microfacet} for incoming light direction $\dirI$ and outgoing view direction $\dirO$ as
\begin{align}
	\brdf = \frac{\fresnel \cdot \geom \cdot \ndf}{4 \cdot \ndotv \cdot \ndotl},
\end{align}
where $\sprod{\cdot}{\cdot}$ denotes the dot product, $\dirN$ is the macroscopic surface normal, $\dirH := \nicefrac{(\dirL+\dirV)}{\|\dirL+\dirV\|}$ is the halfway vector, $\fresnel$ is the Fresnel reflectance term for perfectly smooth microfacets, $\geom$ is the geometric shadowing and masking term, and $\ndf$ is the normal distribution function (NDF) parameterized by a roughness value $\alpha$ which we omit for brevity.
Since we are assuming perfectly smooth microfacets, the halfway vector $\dirH$ is identical to the microfacet normal $\dirM$ that reflects light from $\dirI$ to $\dirO$.
In contrast to previous work \cite{zirr2016real}, we do not require a multi-scale formulation where the reflectance behavior of individual microfacets is again defined by even smaller microfacets.
Starting from the continuous NDF $\ndf$ with an infinite number of microfacets, we consider a finite number of smooth microfacets in the following.
%


\begin{figure}
	\centering
	\includegraphics[width=0.9\linewidth]{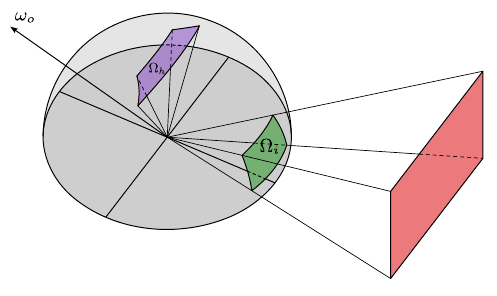}
	\caption{
		The light source (red) defines a spherical polygon $\domainI$ (green) that is transformed to the halfway vectors $\domainH$ (purple) over which the continuous NDF is integrated.
	}
	\label{fig:hemisphere}
\end{figure}

\paragraph{Rendering Equation}
The BRDF is used in the rendering equation\cite{kajiya1986rendering} to compute the outgoing radiance towards the observer.
\begin{align}
	\label{eq:rendering}
	\Lo &= \int_\domainHemi \brdf \cdot \ndotl \cdot \Li \dx{\dirL} \\
	\label{eq:rendering_const_l}
	&= \totalLi \cdot \int_\domainL \brdfnol \dx{\dirL},
\end{align}
where $\domainHemi$ is the upper hemisphere around $\dirN$.
We restrict the integration domain to the incoming light directions $\domainI \subseteq \domainHemi$, defined by a single area light.
The corresponding domain of halfway vectors $\domainH \subseteq \domainHemi$ are exactly those that reflect any $\dirI \in \domainI$ to $\dirO$.
The geometric configuration is shown in \cref{fig:hemisphere}.
The light source diffusely emits a constant amount of radiance $\totalLi$ in all directions.
This allows us to factor out the incoming radiance $\totalLi$.
%
%
We further define $\brdfnol := \brdf \cdot \ndotl$ for brevity.
In the following, we derive an approximation of the outgoing radiance for microfacet BRDFs with a discrete normal distribution illuminated by area lights.
\paragraph{Mean Reflectance}
The BRDF is assembled from multiple terms that all depend on the incoming and outgoing light directions.
Given the incoming directions $\domainI$ and outgoing direction $\dirV$, we assume that all microfacets either reflect an identical amount of light per microfacet area, or none at all due to incorrect orientation.
%
%
We treat individual microfacets as identically sized, irrespective of their orientation, therefore $\ndf$ is directly proportional to the \emph{number} of microfacets and we are not concerned with their projected area.
Starting from \cref{eq:rendering_const_l}, a multiplication of $\Lo$ with one yields
\begin{align}
	\Lo
	= \totalLi \cdot \underbrace{\frac{\int_\domainI \brdfnol \dx{\dirI}}{\intDLong}}_{=: \meanRefl} \cdot \intDLong
	\label{eq:mean_refl_def_2}
\end{align}
Changing variables from $\domainI$ to $\domainH$ in the numerator induces an additional factor of $\left|\nicederivative{\dirI}{\dirH}\right| = 4\cdot\hdotl$.
This lets us interpret $\meanRefl$ as the \emph{mean} reflected light per microfacet area from the incoming directions $\domainI$ to the outgoing direction $\dirO$
\begin{align}
	\meanRefl
	= \int_\domainH \frac{\fresnel \cdot \geom \cdot \hdotl}{\ndotv} \cdot \frac{\ndf}{\int_\domainH \ndfArgs{\dirH'} \dx{\dirH'}} \dx{\dirH}.
\end{align}
\paragraph{Discrete NDF}
Traditionally, an infinite number of microfacets is assumed to contribute to the BRDF.
When modeling glinty appearance, a finite number of microfacets $\Nfoot$ depending on the pixel footprint $\pxFoot$ visible from the observer is used to determine the amount of reflected light.
Since we are not going to explicitly instantiate these microfacets and only count them stochastically, we assume that the spatial and angular distributions of microfacets are independent.
The pixel footprint $\pxFoot$ only affects the number of candidate microfacets $\Nfoot$, not the distribution of their orientations.
%
%
Conversely, the light source only affects the probability for a microfacet to be oriented correctly.
The continuous NDF $\ndf$ measures the microfacet area with orientation $\dirH$ per macroscopic unit surface area.
We first define the total microfacet area $\intDHemi := \intDHemiLong$ as the integral over all orientations.
%
The discrete NDF $\ndfFoot$ with $\Nfoot$ microfacets is then defined as
\begin{align}
	\ndfFoot := \intDHemi \cdot \frac{1}{\Nfoot} \cdot \sum_{k=1}^{\Nfoot} \delta_\dirH\left(\dirH - \dirHk\right),
	\label{eq:discrete_ndf}
\end{align}
%
where $\delta_\dirH(\cdot)$ is the Dirac-delta function with respect to the solid angle measure of $\dirH$.
The individual discrete microfacets $\dirHk$ are distributed proportionally to the continuous NDF $\dirHk \propto \nicefrac{\ndfArgs{\dirHk}}{\intDHemi}$ for $1 \leq k \leq \Nfoot$.
The discrete microfacets are assumed to be identically sized, irrespective of their orientation.
%
In order to preserve the total microfacet area of the continuous NDF, a factor of $\intDHemi$ is required in the discrete NDF.
This distributes the total microfacet area equally among the $\Nfoot$ discrete microfacets.
%
%
%
For all measurable subsets $A \subseteq \domainHemi$, the construction in \cref{eq:discrete_ndf} yields the equality
%
%
%
\begin{align}
		\int_A \Ex{\ndfFoot} \dx{\dirH} = \int_A \ndf \dx{\dirH},
	\label{eq:ndf_convergence}
\end{align}
which implies equality of the integrands almost everywhere.
We provide a proof of \cref{eq:ndf_convergence} in \cref{app:ndf_convergence}.
\paragraph{Replacing the NDF}
We replace the continuous $\ndf$ in the BRDF model by its discrete counterpart $\ndfFoot$, and define the outgoing radiance $\LoFoot$ of the modified model.
Given that a microfacet $\dirHk$ reflects light from the source to the observer, we assume an identical amount of reflection for a given light source and observer.
Because of \cref{eq:ndf_convergence}, we use the mean reflectance $\meanRefl$ of the continuous BRDF model in our approximation of the discrete version.
We replace the continuous NDF in \cref{eq:mean_refl_def_2} with its discrete version (\cref{eq:discrete_ndf}) to get an approximation for the outgoing radiance
\begin{align}
	\LoFoot
	&= \totalLi \cdot \int_\domainI \frac{\fresnel\cdot\geom\cdot\ndfFoot}{4\cdot\ndotv} \dx{\dirI} \\
	&\approx \totalLi \cdot \meanRefl \cdot \int_\domainH \ndfFoot \dx{\dirH},
	\label{eq:lo_fg_discrete}
\end{align}
where for $\Nfoot \rightarrow \infty$ equality is reached again.
The problem is thus reduced to estimating the integral over the discrete NDF $\ndfFoot$.
\paragraph{Integrating the Discrete NDF}
The discrete nature of our NDF induces the following problem:
Any distribution of finitely many microfacets will always evaluate $\ndfFoot = 0$ for externally determined directions $\dirH$ (e.g. not sampled from $\ndfFoot$ itself but from a light source).
Therefore, we need to relax the problem of evaluating the BRDF for discrete light directions, and consider the integral over all possible microfacet orientations that reflect light from $\domainI$ to $\dirO$
\begin{align}
	\int_\domainH \ndfFoot \dx{\dirH} = \frac{\intDHemi}{\Nfoot} \cdot \sum_{k=1}^{\Nfoot} \mathbf{1}_\domainH\left(\dirHk\right) \dx{\dirH},
\end{align}
where $\mathbf{1}_\domainH(\dirH)$ is the characteristic function of $\domainH$ evaluating to $1$ if $\dirH \in \domainH$ and $0$ otherwise.
Traversing the microfacets in the pixel footprint \cite{jakob2014discrete} is too expensive for real-time applications.
Therefore, previous work \cite{zirr2016real,deliot2023real} employs a counting model based on the binomial distribution with $\Nfoot$ trials and some success probability $\probH$, which is defined in the next paragraph:
\begin{align}
	\label{eq:binomial_intd}
	\int_\domainH \ndfFoot \dx{\dirH} = \intDHemi \cdot \frac{b(\Nfoot, \probH)}{\Nfoot}.
\end{align}
With respect to the previous work we build upon, our contribution focuses on choosing $\probH$, taking into account the light source size.
\paragraph{Reflection Probability}
Integrating the discrete NDF means counting the number of microfacets that reflect light from the $\domainI$ to $\dirO$, i.e. we need to count the number of microfacets $\dirHk \in \domainH$.
Using the counting model based on the binomial distribution, we now define the probability $\probH$ for a single microfacet $\dirHk$ to be correctly oriented,
for which only a simple heuristic was used in previous work \cite{deliot2023real}.
%
We assume that the microfacets are identically sized, irrespective of their orientation.
The continuous NDF $\ndf$ measures the microfacet surface area with orientation $\dirH$ per macroscopic unit surface area.
The discrete probability $\probH$ for any specific microfacet $\dirHk$ to reflect light from the source to the observer is proportional to the integrated continuous NDF and needs to be normalized by the total microfacet area over the whole hemisphere $\domainHemi$ of possible microfacet orientations:
\begin{align}
	\tikzboxed[thick,rounded corners=2pt]{\probH := \frac{\intDLong}{\intDHemiLong} = \frac{\intD}{\intDHemi}}\;,
	\label{eq:probh_def}
\end{align}
where we define the integrated NDF $\intD := \intDLong$, analogous to the total microfacet area $\intDHemi$.
%

%
The integration domain is not directly defined on the microfacet orientations $\dirH \in \domainH$, but by the light source over incoming directions $\dirL \in \domainL$.
We perform a change of variables from $\domainH$ to $\domainI$, which introduces an additional factor of $\left|\derivative{\dirH}{\dirI}\right| = \nicefrac{1}{4\hdotl}$
\begin{align}
	\intD = \intDLong = \int_\domainL \frac{\ndf}{4\hdotl} \dx{\dirL}.
	\label{eq:total_microfacet_area}
\end{align}
Fortunately, closed-form expressions of the total microfacet area $\intDHemi$ exist for common distributions, presented in \cref{app:total_microfacet_area}.

\begin{figure}
	\centering
	\def\placeholdersize{0.15\linewidth}
	\def\plotsize{1.9cm}
	\def\smallplotsize{0.95cm}
	\newcommand{\imgpath}[1]{images/lobes/lobes_#1.png}
	\begin{tikzpicture}[node distance=.5em,every node/.style={inner sep=0,outer sep=0}, remember picture]
\pgfplotsset{
	lobeplot/.style={
		clip=true,
		unbounded coords=jump,
		axis on top=false,
		xmin=0,xmax=360, 
		ymin=0,ymax=1, 
		width=\plotsize,height=\plotsize,
		scale only axis,
		xtick=\empty,
		ytick=\empty,
		xticklabel=\empty,
		yticklabel=\empty,
		tick label style={font=\scriptsize,inner sep=1,outer sep=1},
		tick style={major tick length=2pt},
		y axis line style={draw=none}
	},
	smalllobeplot/.style={
		clip=true,
		unbounded coords=jump,
		axis on top=false,
		xmin=0,xmax=360, 
		ymin=0,ymax=1, 
		width=\smallplotsize,height=\smallplotsize,
		scale only axis,
		xtick=\empty,
		ytick=\empty,
		xticklabel=\empty,
		yticklabel=\empty,
		tick label style={font=\scriptsize,inner sep=1,outer sep=1},
		tick style={major tick length=2pt},
		y axis line style={draw=none}
		},
	dirplot/.style={
		only marks, mark=x, white,
		every node/.style={left=0}
	},
	plot graphics/lobeimgfull/.style={
		xmin=45,ymin=-1.414,xmax=45,ymax=1.414
	},
	plot graphics/lobeimgupper/.style={
		xmin=0,ymin=-1,xmax=45,ymax=1.414,includegraphics={trim=0 64 0 0,clip}
	},
	plot graphics/lobeimglower/.style={
		xmin=45,ymin=-1.414,xmax=0,ymax=1,includegraphics={trim=0 0 0 64,clip}
	}
}
\newcommand{\makeplot}[7][]{
	\node[#1, minimum size=\plotsize] (#2) {};
	\begin{polaraxis}[at={(#2)},anchor=center,lobeplot]
		\addplot graphics[lobeimgfull] {\imgpath{#3#4}};
		\addplot[dirplot] coordinates {(0, #7)} node {\scriptsize $\dirV$};
	\end{polaraxis}
	\ifx&#5&\else
	\node[above right=of #2.center, minimum size=\smallplotsize] (upper_#2) {};
	\begin{polaraxis}[at={(upper_#2)},anchor=center,smalllobeplot]
		\addplot graphics[lobeimgfull] {\imgpath{#3#5}};
	\end{polaraxis}
	\fi
	\ifx&#6&\else
	\node[below right=of #2.center, minimum size=\smallplotsize] (lower_#2) {};
	\begin{polaraxis}[at={(lower_#2)},anchor=center,smalllobeplot]
		\addplot graphics[lobeimgfull] {\imgpath{#3#6}};
	\end{polaraxis}
	\fi
	\expandafter\xdef\csname prevplot\endcsname{#2}
}
\newcommand{\makerow}[5][]{
	\makeplot[#1]
		{#2_brdf}
		{#3}{brdf}
		{diff_ndf_brdf}{}
		{#4}

	\makeplot[right=of \prevplot]
		{#2_brdf_ltc}
		{#3}{brdf_ltc}
		{diff_ndf_brdf_ltc}{diff_brdf_brdf_ltc}
		{#4}

	\makeplot[right=of \prevplot]
		{#2_ndf}
		{#3}{ndf}
		{}{}
		{#4}

	\makeplot[right=of \prevplot]
		{#2_ndf_ltc}
		{#3}{ndf_ltc}
		{diff_ndf_ndf_ltc}{}
		{#4}

	\node[above=of #2_brdf] {#5};

	\draw[line width=0.6pt] ($(#2_brdf.north west)+(0,1.4em)$) -- ($(#2_ndf_ltc.north east)+(0,1.4em)$);

	\expandafter\xdef\csname prevrow\endcsname{#2_brdf}
}

\makerow[at={(0,0)}]
	{row_1}
	{thetav_30_sqrt_alpha_0.5_}{0.5}
	{\tiny $\sqrt\alpha = 0.5$, $\theta_o=\ang{30}$}

\makerow[below=2em of \prevrow]
	{row_2}
	{thetav_75_sqrt_alpha_0.5_}{0.966}
	{\tiny $\sqrt\alpha = 0.5$, $\theta_o=\ang{75}$}

\makerow[below=2em of \prevrow]
	{row_3}
	{thetav_30_sqrt_alpha_0.8_}{0.5}
	{\tiny $\sqrt\alpha = 0.8$, $\theta_o=\ang{30}$}

\makerow[below=2em of \prevrow]
	{row_4}
	{thetav_75_sqrt_alpha_0.8_}{0.966}
	{\tiny $\sqrt\alpha = 0.8$, $\theta_o=\ang{75}$}

\node[above=1.8em of row_1_brdf] {\small BRDF};
\node[above=1.8em of row_1_ndf] {\small NDF};
\node[above=1.8em of row_1_brdf_ltc] {\small BRDF LTC};
\node[above=1.8em of row_1_ndf_ltc] {\small NDF LTC};

	\end{tikzpicture}
	\caption{
		Comparison of lobe shapes defined by the continuous BRDF and NDF, as well as their LTC approximations for $\sqrt{\alpha} \in \{0.5,0.8\}$ and $\theta_o \in \{\ang{30},\ang{75}\}$ projected into the tangent plane.
		The top difference images of the BRDF, BRDF LTC and NDF LTC columns show the signed difference with the NDF.
		The bottom difference image of BRDF LTC shows the signed difference with the BRDF.
	}
	\label{fig:lobes}
\end{figure}
\paragraph{Computing the Outgoing Radiance}
Plugging the integration for the discrete NDF $\ndfFoot$ from \cref{eq:binomial_intd} into \cref{eq:lo_fg_discrete}, we write the reflected radiance as
%
\begin{align}
	\LoFoot \approx \totalLi \cdot \mathcal{FG} \cdot \frac{\intDHemi \cdot b(\Nfoot, \probH)}{\Nfoot}.
\end{align}
Applying the definition of $\meanRefl$ (\cref{eq:mean_refl_def_2}) and $\probH$ (\cref{eq:probh_def}) expands the reflected radiance of the BRDF with discrete NDF to
\begin{align}
	\LoFoot &\approx \totalLi \cdot \int_\domainI \brdfnol \dx{\dirI} \cdot \frac{\intDHemi}{\intD} \cdot \frac{b(\Nfoot, \probH)}{\Nfoot}
	\\
	\label{eq:lo_fin}
	&= \tikzboxed[thick,rounded corners=2pt]{\Lo \hide{\int_\domainI \brdf\cdot\ndotl \dx{\dirI}} \cdot \frac{b(\Nfoot, \probH)}{\Nfoot \cdot \probH}} \;.
\end{align}
Notice that the first integral is simply the evaluation of the continuous BRDF model.
This is then simply modulated by an evaluation of the binomial distribution divided by its expected value.
The latter part is responsible for taking the difference between the continuous and discrete versions of the model into account.
 
\begin{figure}
	\centering
	\def\placeholdersize{0.15\linewidth}
	\def\plotsize{2.0cm}
	\subfloat[BRDF $\FGD$]{
		\begin{tikzpicture}[node distance=.5em,every node/.style={inner sep=0,outer sep=0}, remember picture]
			\node[at={(0,0)}] (plot) {};
			\begin{axis}[
				at={(plot)},anchor=north,
				clip=true,
				y dir=reverse,
				unbounded coords=jump,
				axis on top=true,
				axis lines=box,
				xmin=0,xmax=1,
				ymin=0,ymax=1,
				width=\plotsize,height=\plotsize,
				scale only axis,
				xtick={0,1},
				ytick={0,1},
				tick label style={font=\scriptsize,inner sep=1,outer sep=1},
				tick style={major tick length=2},
				tick align=outside,
				xtick pos=left,
				ytick pos=left,
				xlabel=\small$\sqrt{\alpha}$,
				ylabel=\small$\sqrt{\ndotv}$,
				every axis x label/.style={inner sep=1,outer sep=1,at={(ticklabel* cs:0.5,0)},anchor=north},
				every axis y label/.style={inner sep=1,outer sep=1,at={(ticklabel* cs:0.5,0)},anchor=south,rotate=90}
				]
				\addplot graphics[xmin=0,ymin=0,xmax=1,ymax=1] {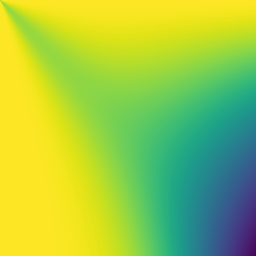};
			\end{axis}
			\node[above=0.2em of plot] (cmap) {};
			\begin{axis}[
				at={(cmap)},anchor=south,
				clip=true,
				unbounded coords=jump,
				axis on top=true,
				axis lines=box,
				xmin=0.3,xmax=1,
				ymin=0,ymax=1,
				width=\plotsize,height=0.15cm,
				scale only axis,
				xtick={0.3,1},
				ytick=\empty,
				tick label style={font=\scriptsize,inner sep=1,outer sep=1},
				tick style={major tick length=2},
				tick align=outside,
				xtick pos=right
				]
				\addplot graphics[ymin=0,xmin=0.3,ymax=1,xmax=1,includegraphics={angle=90}] {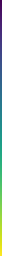};
			\end{axis}
		\end{tikzpicture}
		\label{fig:fgd:brdf}%
	}
	\subfloat[NDF $\ndfFGD$]{
		\begin{tikzpicture}[node distance=.5em,every node/.style={inner sep=0,outer sep=0}]
			\node[at={(0,0)}] (plot) {};
			\begin{axis}[
				at={(plot)},anchor=north,
				clip=true,
				y dir=reverse,
				unbounded coords=jump,
				axis on top=true,
				axis lines=box,
				xmin=0,xmax=1,
				ymin=0,ymax=1,
				width=\plotsize,height=\plotsize,
				scale only axis,
				xtick={0,1},
				ytick={0,1},
				tick label style={font=\scriptsize,inner sep=1,outer sep=1},
				tick style={major tick length=2},
				tick align=outside,
				xtick pos=left,
				ytick pos=left,
				xlabel=\small$\sqrt{\alpha}$,
				ylabel=\small$\sqrt{\ndotv}$,
				every axis x label/.style={inner sep=1,outer sep=1,at={(ticklabel* cs:0.5,0)},anchor=north},
				every axis y label/.style={inner sep=1,outer sep=1,at={(ticklabel* cs:0.5,0)},anchor=south,rotate=90}
				]
				\addplot graphics[xmin=0,ymin=0,xmax=1,ymax=1] {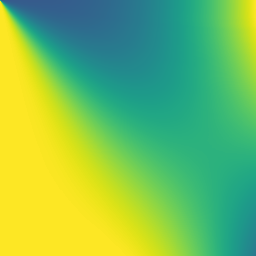};
			\end{axis}
			\node[above=0.2em of plot] (cmap) {};
			\begin{axis}[
				at={(cmap)},anchor=south,
				clip=true,
				unbounded coords=jump,
				axis on top=true,
				axis lines=box,
				xmin=0.3,xmax=1,
				ymin=0,ymax=1,
				width=\plotsize,height=0.15cm,
				scale only axis,
				xtick={0.3,1},
				ytick=\empty,
				tick label style={font=\scriptsize,inner sep=1,outer sep=1},
				tick style={major tick length=2},
				tick align=outside,
				xtick pos=right
				]
				\addplot graphics[ymin=0,xmin=0.3,ymax=1,xmax=1,includegraphics={angle=90}] {images/fgd/color_map_viridis.png};
			\end{axis}
		\end{tikzpicture}
		\label{fig:fgd:ndf}%
	}
	\subfloat[Difference]{
		\begin{tikzpicture}[node distance=.5em,every node/.style={inner sep=0,outer sep=0}]
			\node[at={(0,0)}] (plot) {};
			\begin{axis}[
				at={(plot)},anchor=north,
				clip=true,
				y dir=reverse,
				unbounded coords=jump,
				axis on top=true,
				axis lines=box,
				xmin=0,xmax=1,
				ymin=0,ymax=1,
				width=\plotsize,height=\plotsize,
				scale only axis,
				xtick={0,1},
				ytick={0,1},
				tick label style={font=\scriptsize,inner sep=1,outer sep=1},
				tick style={major tick length=2},
				tick align=outside,
				xtick pos=left,
				ytick pos=left,
				xlabel=\small$\sqrt{\alpha}$,
				ylabel=\small$\sqrt{\ndotv}$,
				every axis x label/.style={inner sep=1,outer sep=1,at={(ticklabel* cs:0.5,0)},anchor=north},
				every axis y label/.style={inner sep=1,outer sep=1,at={(ticklabel* cs:0.5,0)},anchor=south,rotate=90}
				]
				\addplot graphics[xmin=0,ymin=0,xmax=1,ymax=1] {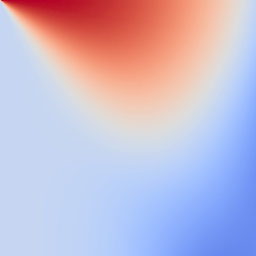};
			\end{axis}
			\node[above=0.2em of plot] (cmap) {};
			\begin{axis}[
				at={(cmap)},anchor=south,
				clip=true,
				unbounded coords=jump,
				axis on top=true,
				axis lines=box,
				xmin=-0.4285,xmax=0.4285,
				ymin=0,ymax=1,
				width=\plotsize,height=0.15cm,
				scale only axis,
				xtick={-0.4,0,0.4},
				ytick=\empty,
				tick label style={font=\scriptsize,inner sep=1,outer sep=1},
				tick style={major tick length=2},
				tick align=outside,
				xtick pos=right
				]
				\addplot graphics[ymin=0,xmin=-0.4285,ymax=1,xmax=0.4285,includegraphics={angle=90}] {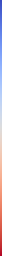};
			\end{axis}
		\end{tikzpicture}
		\label{fig:fgd:diff}%
	}
	\caption{
		Given viewing direction $\dirO$ and roughness $\alpha$, (\subref{fig:fgd:brdf}) we plot the directional albedo $\FGD$ of the BRDF for $\fresnel=1$ and (\subref{fig:fgd:ndf}) the potentially reflecting microfacet area $\ndfFGD$.
		%
		(\subref{fig:fgd:diff}) We plot the difference after scaling $\FGD$ to minimize the squared error with $\ndfFGD$.
	}
	\label{fig:fgd}
\end{figure}
\paragraph{Differentiation from Previous Work}
While Jakob et al. \cite{jakob2014discrete} propose a model where each microfacet is implicitly defined, but can be exactly located by traversing a hierarchical data structure, the approaches we build upon use a purely stochastic representation of the microfacet orientations \cite{zirr2016real,deliot2023real}.
We only have the number of microfacets in the pixel footprint, the probability of correct orientation and a random seed as input.
Our predecessors \cite{zirr2016real,deliot2023real} determine the probability that an individual microfacet is oriented correctly based on a heuristic, where the continuous microfacet distribution $\ndf$ was used directly as a proxy for the discrete probability of correct orientation.
Since the continuous probability density functions can attain values $\gg 1$, they simply divided by the maximum value of the continuous distribution $\ndf$ to attain $\probH = \nicefrac{(R\cdot \ndf)}{\ndfArgs{\dirN}} \in [0, 1]$ for point light sources, where $R \in [0,1]$ is a user-defined parameter that allows for manual tuning of the model.
While this definition delivers a valid probability parameter for the binomial distribution, this quantity ignores the size of the light source, which can be compensated by manually tuning the $R$ parameter.
Our contribution lies in a definition of the probability $\probH$, that intrinsically takes the light source size into account.
We utilize this insight to find a real-time approximation to handle spatially constant diffuse area lights (\cref{sec:area_lights}).
In addition, we formulate an alternative definition of $\probH$ for point and directional lights that does not rely on the $R$ parameter (\cref{sec:infinitesimal_lights}), and can directly be employed in existing implementations \cite{deliot2023real}.
%

%
\section{Area Lights Illumination}
\label{sec:area_lights}

In the previous section, we derived a formulation (\cref{eq:lo_fin}) for the outgoing radiance of microfacet BRDF models with a discrete normal distribution $\LoFoot$ for an arbitrary domain of incoming light directions $\domainI$.
This section applies the formulation to real-time area lights with constant diffuse emission, for which the outgoing radiance is typically approximated using LTCs \cite{heitz2016real}.
\paragraph{LTC Distribution}
We approximate the outgoing radiance $\Lo$ for the continuous microfacet BRDF using LTCs \cite{heitz2016real}.
In order to determine $\probH$, we need to integrate the continuous NDF $\intDLong$ without additional factors like in the BRDF.
It might seem plausible to simply apply the same idea underlying LTCs since $\ndf$ is a linearly transformed cosine by definition, at least for GGX distributions\cite{heitz2018sampling}.
The integration domain of the light source however is not defined in the domain of halfway vectors, and a transformation from $\dirI$ to $\dirH$ does not preserve straight lines, which is a requirement for LTC integration over spherical polygons.
The NDF $\ndf$ is similarly shaped to the corresponding microfacet BRDF $\brdf$ when transformed to the light direction $\dirI$, though not identical.
%
%
A comparison of the NDF and BRDF with their LTC approximations is shown in \cref{fig:lobes}.
In order to reuse the result of the LTC integration at runtime, we propose to approximate $\intD$ using the \emph{same} LTC lobe as the outgoing radiance of the continuous microfacet BRDF $\Lo$, despite the relatively large difference between the BRDF LTC approximation and the target NDF.
Fitting a separate LTC lobe that resembles the NDF more closely is possible, but our experiments have shown that this does not lead to a meaningful difference in visual quality.
Approximating the NDF using the same LTC lobe as the BRDF has an interesting implication:
It makes our technique invariant to the size of the area light, i.e. using one large light source is as good as using many small light sources, which we prove in \cref{app:subdivision}.
Reusing the same LTC lobe has the additional advantage, that the support of both LTC approximations is identical, i.e.
\begin{align}
	\brdf \neq 0 \qquad \Leftrightarrow \qquad \ndf \neq 0.
\end{align}
\paragraph{Directional Albedo}
The directional albedo of the BRDF given $\dirO$, also called $\FGD$ is defined as the integral over the hemisphere
\begin{align}
	\FGD = \int_\domainHemi \brdf \cdot \ndotl \dx{\dirI}.
\end{align}
A similar quantity can be defined for the NDF that measures the area of microfacets that are visible from the observer\cite{heitz2018sampling} and also result in a valid reflection $\dirI \in \domainHemi$
\begin{align}
	\ndfFGD = \int_\domainHemi \frac{\ndf}{4\cdot\hdotl} \dx{\dirI}.
\end{align}
Note that this is different from the total microfacet area $\intDHemi$ since the hemisphere of light directions only maps to a subset of microfacet orientations.
The difference between the directional albedo $\FGD$ and the potentially reflecting microfacet area $\ndfFGD$ is shown in \cref{fig:fgd}.
\paragraph{LTC Scale}
For the shape of the LTC lobe, we argued above that there is no meaningful difference between the BRDF approximation and the target NDF lobe.
For the scale of the LTC lobe, we argue that the difference is very significant (c.f. \cref{fig:fgd}).
If the proper scale is not taken into account, we might substantially overestimate $\probH$, maybe resulting in invalid $\probH > 1$.
Since LTC distributions naturally integrate to $1$, we simply scale them with $\ndfFGD$.
Given $\alpha$ and $\ndotv$, $\ndfFGD$ is computed ahead of time and stored together with the $\FGD$ for the continuous BRDF in a 2D table.
%

%
\section{Infinitesimal Light Sources}
\label{sec:infinitesimal_lights}
So far our goal was to enable rendering of glinty materials illuminated by area lights.
The theory presented in the previous chapters does not allow for infinitesimally small light sources.
%
%
Since the light direction $\dirI$ and the orientation of a discrete microfacet $\dirHk$ are independent,
the product of the corresponding Dirac-delta functions will always evaluate to zero, and therefore yield $\probH = 0$.
In other words: the numerator in \cref{eq:probh_def} is zero for $\domainI = \{\dirI\}$.
However, glints are usually the strongest under very small and bright light sources, e.g. the sun which is usually modeled as a directional light source.
The area light assumption has implications for light sources that are typically assumed to be infinitesimally small as well.
We handle light sources that are traditionally considered infinitesimally small by defining a small size for point light sources, treating them as little balls, or an opening angle for directional light sources.
The choice of this parameter will have a direct impact on the appearance of the glinty material.
To make them compatible with our framework, infinitesimally small light sources are replaced by a finite spherical cap of incoming directions with solid angle area $\srof{\domainI} =2\pi(1-\cos\gamma) \,[\unit{\steradian}]$, where $\gamma$ is the half-angle of the corresponding cone.
We assume $\ndf$ constant and approximate $\intD \approx \srof{\domainI} \cdot \frac{\ndf}{4\cdot\hdotl}$.
If the spherical cap associated with a point light is sufficiently small, the resulting error is also small.
Smooth surfaces with low roughness $\alpha$ require a smaller $\gamma$ due to the increasingly sharp peak in the continuous NDF.
For point light sources, the probability $\probH$ is thus given by
\begin{align}
	\probH = \frac{\intDLong}{\intDHemiLong}
	\approx \frac{\srof{\domainI} \cdot D(\dirH)}{\intDHemi \cdot 4\cdot\hdotl}.
\end{align}
Assuming $\intD \approx \srof{\domainI} \cdot \frac{\ndf}{4\cdot\hdotl}$ is reasonable only when the microfacet surface is sufficiently rough and the light source sufficiently small.
Otherwise the approximation might result in $\probH > 1$.
Should that happen, we suppose that clamping $\probH \in [0, 1]$ is sufficient, replacing the glinty appearance with the continuous BRDF in these situations, usually at the center of a specular highlight.
Alternatively, the light source could be modeled as a small area light.
%

%

\begin{table}
	\centering
	\newcommand{\tablepath}[2]{images/runtime/table_#1_flat_#2.csv}
	\setlength\tabcolsep{5.5pt}
	\begin{tabular}{ccccc}
		\toprule
		Light & Smooth & \citeq{deliot2023real} & Ours & + NDF LTC \\

		\midrule
		&\multicolumn{4}{l}{\tiny RTX 4090 @ $3840\times2160$ \hfill [\unit{\milli\second}]} \\
		\midrule
		\csvreader[head to column names,late after line=\\, late after last line=\\]{\tablepath{4k}{rtx4090}}{}
			{ \light & $\smoothE \pm \smoothD$ & $\simpleE \pm \simpleD$ & $\oursE \pm \oursD$ & $\ndfltcE \pm \ndfltcD$ }

		\midrule
		&\multicolumn{4}{l}{\tiny RTX 4070M @ $3840\times2160$ \hfill [\unit{\milli\second}]} \\
		\midrule
		\csvreader[head to column names,late after line=\\, late after last line=\\]{\tablepath{4k}{rtx4070m}}{}
			{ \light & $\smoothE \pm \smoothD$ & $\simpleE \pm \simpleD$ & $\oursE \pm \oursD$ & $\ndfltcE \pm \ndfltcD$ }



		\bottomrule
	\end{tabular}
	\caption{
		Runtime per frame and standard deviation on two GPUs.
		We render a full screen plane illuminated by one light source.
		The first rows shows the baseline overhead per frame without meaningful shading.
		The second and third rows use a directional and a rectangular area light source, respectively.
		The columns show the continuous BRDF without glints (Smooth),
		the previous method (\citeq{deliot2023real}) for which we implement area lights by evaluating glints for the mean light direction only,
		as well as our method reusing the BRDF LTC lobe (Ours) and using a separately optimized LTC lobe for the NDF integration (+NDF LTC).
		%
		%
	}
	\label{tab:runtime}
\end{table}

\input{source/figure/microfacet_roughness.tex}
The integral in the outgoing radiance of the continuous BRDF model $\Lo$ with an underlying Dirac-delta distribution of the point light source reduces to an evaluation of the integrand for the $\dirI$ associated with the light source.
The outgoing radiance for the BRDF with discrete NDF \cref{eq:lo_fin} is given by
\begin{align}
	\LoFoot &\approx \totalLi \cdot \int_\domainI \brdfnol \dx{\dirI} \cdot \frac{b(\Nfoot, \probH)}{\Nfoot \cdot \probH} \\
	\approx \totalLi \cdot &\frac{\fresnel \cdot \geom}{\ndotv} \cdot \frac{ \intDHemi \cdot \hdotl}{\srof{\domainI}} \cdot \frac{b(\Nfoot, \probH)}{\Nfoot}.
\end{align}
The spherical cap of incoming light directions $\domainI$ already makes an occurrence in previous work \cite{jakob2014discrete}, where a small spherical cap with half-angle $\gamma \in [\ang{0.5}, \ang{6}]$ is assigned to each ray direction $\dirI$.
This is necessary to facilitate the counting of microfacets that reflect light form $\domainI$ to $\dirO$, which requires $\srof{\domainI} > 0$.
In principle, this is very similar to our setup for infinitesimally small lights.
However, they specify $\gamma$ globally, irrespective of the underlying light source.
Whereas in our framework, $\gamma$ is a tunable parameter of individual point-like light sources.
%

%
\section{Evaluation}
\label{sec:results}
We implemented\footnote{Unity3D project available on GitHub at \url{https://github.com/tomix1024/AreaLightGlintsUnityProject}} our method based on the publicly available source code of Deliot and Belcour\cite{deliot2023real} in the High Definition Render Pipeline 14.0.10 for Unity3D 2022.3.
We restrict our implementation to the isotropic GGX microfacet BRDF with the Smith shadowing assumption\cite{heitz2014understanding} and Schlick's approximation\cite{schlick1994inexpensive} for the Fresnel term.
The performance of our method is validated on an Nvidia RTX 4090 GPU and a mobile Nvidia RTX 4070 GPU power-limited to $\SI{35}{\watt}$ on Windows 11 with Unity's DirectX 12 backend.
The measured runtimes are reported in \cref{tab:runtime}.
We inherit the good performance characteristics of the employed building blocks \cite{heitz2016real,deliot2023real}, and combining them only adds a small additional overhead.
For directional light sources, we measure a slight improvement in runtime with our method.
%
%
%
Since the only difference between the baseline and our method for directional lights is in the computation of $\probH$, which should be of similar complexity,
the difference in runtime is likely due to a difference in code generation.
\input{source/figure/area_light_ref_ablation.tex}
We reproduce the appearance of the baseline \cite{deliot2023real} for directional lights.
It exposes an ad-hoc parameter $R$ to the user that acts as a linear attenuator on their $\probH = \nicefrac{(R \cdot \ndf)}{\ndfArgs{\dirN}}$.
Given the surface roughness and a light source size, we fit the $R$ parameter in \cref{fig:microfacet_roughness} to match our model at normal incidence, ensuring equivalence between the methods in this particular situation.
The baseline can reproduce our appearance, however this requires manually adjusting $R$ depending on both surface roughness and light size.
Our method thus eliminate this additional user parameter from the baseline and replaces it with a physical parameter for the light opening angle.
\cref{fig:microfacet_roughness} also shows a configuration where $\probH > 1$ for our method and $R > 1$ for the baseline, which can easily be worked around by clipping $\probH \in [0, 1]$.
Notice that this configuration requires $R > 1$ in the baseline.
\cref{fig:area_light_ref_ablation} shows the error introduced when approximating the Monte Carlo integrals for $\Lo$ and $\intD$ using LTCs.
Replacing the integral over the BRDF leads to the most notable visual difference, both for the continuous BRDF and our glint BRDF.
Beyond that, approximating the integrated NDF $\intD$ with the \emph{same} LTC as the outgoing radiance of the continuous BRDF $\Lo$ has a minimal impact on the appearance.
Only approximating $\intD$ but not the continuous $\Lo$ leads to more intense glints close to the top edge of the specular highlight since $\probH$ approaches $0$ faster than the continuous $\brdf$, i.e. fewer glints reflecting more light.
In \cref{fig:area_light_ltc_ablation} we show the error introduced by reusing the LTC approximation of the continuous BRDF lobe in comparison to an optimized LTC lobe for the integrated NDF $\intD$.
The distribution of the glints is slightly different in each case.
Our choice for reusing the existing LTC approximation is supported by an easier implementation and negligible change in appearance.
\input{source/figure/area_light_ltc_ablation.tex}
In the beginning, we made the assumption that all microfacets reflect an identical amount of light, irrespective of their orientation.
In \cref{fig:area_light_subdivision} we investigate the error introduced by that assumption.
To simulate a reference, the light source is subdivided into $2^4\times2^4$ smaller light sources for which the continuous BRDF are evaluated separately.
To evaluate the number of microfacets falling into each small light source, we first compute the number of microfacets falling into the whole light source, and then distributing them according to a multinomial distribution into the smaller light source, which works since they partition the incoming light directions and a microfacet can only reflect light from one of the smaller light sources.
In \cref{app:subdivision} we prove that our model is unaffected by subdivision.
It does affect the evaluation of the ground-truth terms though.
%
%

%
We demonstrate the convergence behavior of our method to the underlying continuous microfacet BRDF for $\Nfoot \rightarrow \infty$ in \cref{fig:convergence}.
In the limit there is no perceptible difference between the discrete model with high $\Nfoot$ and the continuous model.
Depending on the surface roughness $\alpha$ and the size of the light source, a different convergence behavior can be observed.
%
%

%
\input{source/figure/area_light_subdivision.tex}
%

%

\begin{figure*}
	\centering
	%
	\def\plotsize{2.2cm}
	\def\convergedGlint{30}
	\def\smoothGlint{0}
	\newcommand{\imgpath}[3]{images/convergence/scale4/#1_glints_#2_#3.png} 
	\newcommand{\smoothpath}[2]{images/convergence/#1_smooth_#2.png} 
	\newcommand{\makecell}[4][]{
		\node [imgnode,#1] (grad_#2) {\includegraphics[height=\plotsize]{\imgpath{#3}{gradient}{#4}}};
		\node [imgnode,right=of grad_#2] (ref_#2) {%
			\includegraphics[height=\plotsize,trim=0 0 256 0,clip]{\imgpath{#3}{\convergedGlint}{#4}}%
			\includegraphics[height=\plotsize,trim=256 0 0 0,clip]{\smoothpath{#3}{#4}}%
		};

		\foreach \i in { 1, ..., 9 } {
			\pgfmathsetmacro{\j}{\i/10}
			\draw[line width=0.5pt,line cap=rect] ($(grad_#2.north west)!\j!(grad_#2.north east)$) -- ($(grad_#2.south west)!\j!(grad_#2.south east)$);
		};

		\draw[line width=0.5pt] (ref_#2.north) -- (ref_#2.south);
		\expandafter\xdef\csname prevcell\endcsname{grad_#2}
	}
	\newcommand{\makecelldesc}[1]{
		\draw[-Latex] ($(grad_#1.north west)+(0.2cm,0.5em)$) -- ($(grad_#1.north east)+(-0.2cm,0.5em)$) node[midway,above=0.2em]{\small increasing $\log\Nfoot$};
		\node [above=0.5em of $(ref_#1.north west)!0.5!(ref_#1.north)$]{\small Glint};
		\node [above=0.5em of $(ref_#1.north east)!0.5!(ref_#1.north)$]{\small Smooth};
	}
	\tikzset{imgnode/.style={
		draw, line width=1pt,line join=round
	}}
	\subfloat[Large Area Light ($5\times5$ at $z=1$)]{%
	\begin{tikzpicture}[node distance=.2em,every node/.style={inner sep=0,outer sep=0}]
		\makecell[]{large_9}{area_light_large}{0.9}
		\makecell[below=of \prevcell]{large_5}{area_light_large}{0.5}
		\makecell[below=of \prevcell]{large_1}{area_light_large}{0.1}
		
		\makecelldesc{large_9}
		
		\node[left=of grad_large_9,anchor=south,rotate=90]{\small$\sqrt{\alpha} = 0.1$};
		\node[left=of grad_large_5,anchor=south,rotate=90]{\small$\sqrt{\alpha} = 0.5$};
		\node[left=of grad_large_1,anchor=south,rotate=90]{\small$\sqrt{\alpha} = 0.9$};
	\end{tikzpicture}
	\label{fig:convergence:large}
	}
	\subfloat[Small Area Light ($0.25\times0.25$ at $z=1$ )]{%
	\begin{tikzpicture}[node distance=.2em,every node/.style={inner sep=0,outer sep=0}]
		\makecell[]{small_9}{area_light_small}{0.9}
		\makecell[below=of \prevcell]{small_5}{area_light_small}{0.5}
		\makecell[below=of \prevcell]{small_1}{area_light_small}{0.1}
		\makecelldesc{small_9}
	\end{tikzpicture}
	\label{fig:convergence:small}
	}
	\subfloat[Directional Light ($\gamma=\ang{1.3}$)]{
	\begin{tikzpicture}[node distance=.2em,every node/.style={inner sep=0,outer sep=0}]
		\makecell[]{dir_9}{directional_light}{0.9}
		\makecell[below=of \prevcell]{dir_5}{directional_light}{0.5}
		\makecell[below=of \prevcell]{dir_1}{directional_light}{0.1}
		\makecelldesc{dir_9}
	\end{tikzpicture}
	\label{fig:convergence:dir}
	}
	\caption{
		We demonstrate the convergence behavior $\Nfoot \rightarrow \infty$ to the smooth BRDF model for perceptual roughness $\sqrt{\alpha} \in \{0.1, 0.5, 0.9\}$ (top to bottom).
		Horizontally, we show (\subref{fig:convergence:large}) a large area light, (\subref{fig:convergence:small}) a small area light and (\subref{fig:convergence:dir}) a directional light.
		The left part shows a sequence of renderings with varying seeds where $\Nfoot$ quadruples between adjacent strips.
		The right part compares the converged glint model with high $\Nfoot$ against the smooth BRDF.
	}
	\label{fig:convergence}
\end{figure*}

%

%
\section{Conclusion}
\label{sec:conclusion}
We have introduced a novel framework for rendering glints under illumination from area lights in real time.
The crucial part is to correctly determine the probability that a microfacet is correctly oriented for reflection.
Starting from a discrete normal distribution, our derivations have led to a multiplicative factor for the continuous version of the BRDF.
Our practical implementation is based on LTC approximations \cite{heitz2016real} and an efficient evaluation of binomial distributions \cite{deliot2023real}.
Combining these approaches adds little additional overhead.
\paragraph{Assumptions}
Throughout our derivation we made several assumptions:
First, we assume that all reflecting microfacets reflect light according to the mean reflectance $\meanRefl$.
Investigating subdivided light sources, we have shown that this produces an error, but it is dominated by the following assumption.
For area lights, we assume that both the continuous BRDF and the continuous NDF can be represented by the same LTC lobe.
In practice, the LTC approximation of the continuous BRDF is reused for the continuous NDF.
Our evaluation shows that the LTC approximation of the continuous BRDF has the most significant impact on the visual appearance.
The difference due to an approximation of the continuous NDF with the same lobe is measurable but visually negligible.
For point lights we instead assume that the continuous NDF is locally constant around the given light direction.
The last assumption lies in the approximation of the total microfacet area $\intDHemi$.
We have not encountered any artifacts that we could trace back to this approximation.
While making all these assumptions, our results (\cref{fig:area_light_ref_ablation,fig:area_light_ltc_ablation,fig:area_light_subdivision}) demonstrate good visual quality at decent performance.
\paragraph{Limitations}
One apparent limitation is that our method might exhibit artifacts if $\probH > 1$.
This condition might arise for directional light sources, for which we assume that the continuous NDF $\ndf$ is locally constant.
When the opening angle of the light source is set too large and the surface roughness too low the assumption breaks down.
An easy workaround for this problem is to clip $\probH \in [0, 1]$, effectively falling back to the continuous BRDF.
The original LTC formulation contains an extension for textured light sources \cite{heitz2016real}.
This is a scenario that we do not consider here.
If the contrast in the emissive texture is strong enough, it might have a significant impact on the integrated NDF $\intD$ and should be taken into account.
Another interesting avenue for future work would be to investigate the interaction between glints and iridescent color shifts \cite{belcour2017practical,kneiphof2019real,liu2023real} under area light illumination.
Our theory should also be applicable to image-based lighting, where prefiltering could be applied to the environment map to estimate the integrated NDF $\intD$.
Another apparent limitation of the presented method is that it does not account for visibility with the light source.
Since the domain of light directions plays an important role in our method, it is not sufficient to simply multiply our solution with a visibility estimate \cite{heitz2018combining}.
Instead, the change of area light size should be taken into account when integrating over the light source.
The limitation is mitigated by clipping the spherical polygon of the area light with that of an occluder \cite{kt2021fast}.
\hide{
At first glance we would expect that the BRDF is basically the NDF attenuated due to the Fresnel term $\fresnel$ and masking/shadowing term $\geom$.
This however turns out not to be the case (\cref{fig:fgd}).
There is one pertinent difference between the two cases:
With the BRDF we are interested in the amount of light reflected, and with the NDF we are interested in the \emph{number} of microfacets.
Depending on the projected area of the microfacets, many more microfacets might be required to achieve the same amount of light transport for different orientations.
}
%

\bibliographystyle{eg-alpha-doi}
\bibliography{AreaLightGlints-bibliography}


\appendix

%
\section{NDF Convergence}
\label{app:ndf_convergence}
In the following, we prove that the expected value of the discrete NDF and the continuous NDF in \cref{eq:ndf_convergence} are equal almost everywhere.
Let $A \subseteq \domainHemi$ be measurable.
\begin{align}
	& \int_A \Ex{\ndfFoot} \dx{\dirH} \\
	=& \int_A
	\frac{\intDHemi}{\Nfoot} \cdot \sum_{k=1}^{\Nfoot} \Ex{\delta_\dirH\left(\dirH - \dirHk\right)} \dx{\dirH} \\
	=& \int_A \frac{\cancel{\intDHemi}}{\Nfoot} \cdot \sum_{k=1}^{\Nfoot} \int_\domainHemi \delta_\dirH(\dirH - \dirHk) \cdot \frac{\ndfArgs{\dirHk}}{\cancel{\intDHemi}} \dx{\dirHk} \dx{\dirH} \\
	=& \int_A \frac{1}{\Nfoot} \cdot \sum_{k=1}^{\Nfoot} \int_\domainHemi \delta_\dirH(\dirH - \dirHk) \cdot \ndfArgs{\dirHk} \dx{\dirHk} \dx{\dirH} \\
	=& \int_A \frac{1}{\Nfoot} \cdot \sum_{k=1}^{\Nfoot} \ndf \dx{\dirH} \\
	=& \int_A \ndf \dx{\dirH},
\end{align}
which proves that the expected value of the discrete NDF is equal to the continuous NDF almost everywhere.
%

%
\section{Total Microfacet Area}
\label{app:total_microfacet_area}
There exist exact closed-form expressions for the total microfacet area $\intDHemi$ (\cref{eq:total_microfacet_area}) for both GGX and Beckmann distributions.
%
%
The expression to compute for the GGX distribution is
\begin{align}
	\int_\domainHemi \ndfSymb_\mathrm{GGX}\left(\dirH\right) \dx{\dirH}
	= 1 + \frac{\alpha^2\cdot\log\left(\frac{1+\sqrt{1-\alpha^2}}{\alpha}\right)}{\sqrt{1-\alpha^2}}.
	\label{eq:total_ndf_ggx}
\end{align}
We found no benefit from using a simpler approximation of \cref{eq:total_ndf_ggx}.
The solution for the Beckmann distribution involves an evaluation of $\mathrm{erfc}(\cdot)$.
Therefore, we fit a quadratic polynomial
\begin{align}
	\int_\domainHemi \ndfSymb_\mathrm{Beckmann}\left(\dirH\right) \dx{\dirH}
	&= 1 + \frac{\sqrt{\pi}}{2} \cdot \alpha \cdot \exp\left(\frac{1}{\alpha^2}\right) \cdot \mathrm{erfc}\left(\frac{1}{\alpha}\right) \\
	&\approx 1 + 0.466\cdot\alpha - 0.091\cdot\alpha^2,
\end{align}
which is faster to compute and delivers sufficient accuracy for our purpose.
Since both expressions only depend on the roughness $\alpha$, the functions could be tabulated into an array ahead of time.
%

%
\section{Area Light Subdivision}
\label{app:subdivision}
This appendix proves that the result of our model does not change upon subdivision of the light source if the integrand of $\intD$ can be assumed proportional to the continuous BRDF
\begin{align}
	c \cdot \frac{\ndf}{4\cdot \hdotl} \approxbutequal \brdf \cdot \ndotl,
\end{align}
for some $c \in \mathbb{R}$.
We make this assumption when approximating both the BRDF and the microfacet distribution with the same LTC lobe in \cref{sec:area_lights}.
For validation, we render images where the area light source was subdivided into multiple smaller light sources in \cref{fig:area_light_subdivision}.
When subdividing a large light source, individual microfacets strictly fall into exactly one of the resulting smaller light sources.
Let $\domainHsup[(k)]$ be the domain of microfacet orientations corresponding to the $k$-th light analogous to $\probH$ in \cref{eq:probh_def}, we have
\begin{align}
	\probHsup[(k)] = \frac{\int_{\domainHsup[(k)]} \ndf \dx{\dirH} }{\intDHemi},
\end{align}
and since the $\domainHsup[(k)]$ are a partition of $\domainH$, we also have
\begin{align}
	\sum_{k=1}^{K} b(\Nfoot, \probHsup[(k)]) = b(\Nfoot, \probH).
\end{align}
Summing the outgoing radiance from all the small light sources (\cref{eq:lo_fin}), we get
\begin{align}
	\frac{\LoFoot}{\totalLi} &= \sum_{k=1}^{K} \int_{\domainIsup[(k)]} \underbrace{\brdfnol}_{\approxbutequal c \cdot \frac{\ndf}{4\cdot \hdotl}} \dx{\dirI} \cdot \frac{b(\Nfoot, \probHsup[(k)])}{\Nfoot \cdot \probHsup[(k)]} \\
	&\approxbutequal c \cdot \sum_{k=1}^{K} \int_{\domainHsup[(k)]} \ndf \dx{\dirH} \cdot \frac{\intDHemi \cdot b(\Nfoot, \probHsup[(k)])}{\int_{\domainHsup[(k)]} \ndf \dx{\dirH} \cdot \Nfoot} \\
	&= c \cdot \intDHemi \cdot \frac{\sum_{k=1}^{K} b(\Nfoot, \probHsup[(k)])}{\Nfoot} \\
	&= c \cdot \frac{\intDLong}{\intDLong} \cdot \intDHemi \cdot \frac{b(\Nfoot, \probH)}{\Nfoot} \\
	&= \int_\domainI \brdfnol \dx{\dirI} \cdot \frac{b(\Nfoot, \probH)}{\Nfoot \cdot \probH},
\end{align}
which is the outgoing radiance of the full light source in our model.
Therefore, the outgoing radiance is invariant under subdivision of the light source if the continuous NDF and BRDF are proportional.
%


\end{document}